# Entropy shows that global warming should cause increased variability in the weather*

by John Michael Williams

P. O. Box 2697

Redwood City, CA 94064

jwill@AstraGate.net

\* Based on *aps1998nov15_001* at the American Physical Society Web site.

# Abstract

Elementary physical reasoning seems to leave it inevitable that global warming would increase the variability of the weather.  The first two terms in an approximation to the global entropy may be used to show that global warming has increased the free energy available to drive the weather, and that the variance of the weather has increased correspondingly.

# Introduction

Hasselmann [1] summarized the evidence that there has been about a .5º C warming of the globe over the past century.   Recent findings seem to confirm this warming beyond reasonable doubt [2].   The question remains open of whether this warming should be attributed to human activity.   Regardless of its cause, we attempt here an understanding of the most obvious effect of a secular warming of the Earth's atmosphere:  Increased variability of the weather.

If the Earth's atmosphere and superficial layers of ocean water could be treated as a closed system, it might be possible to quantify the observed temperature rise as an effect of a complex, deterministic collection of closely coupled (*viz*, poorly separated) weather processes.

However, the system involved is an open one driven mainly by a continual influx of radiation from the Sun, and by the rotation of the Earth.   The system is too big to solve deterministically because of:  (*a*) the number of data required to describe its state; (*b*) the necessarily incomplete instrumentation for monitoring its state; (*c*) the difficulty of providing input for such monitoring data, were a computer programmed for prediction; (*d*) the lack of an obvious way of separating the variables underlying the data; (*e*) the lack of a valid way of spatially partitioning the system for long-term analysis; and (*f*) a dearth of accurate historical data much before 1900.

One would have to predict at least air, sea, and land temperature; humidity; local air pressure, clouds, wind, and precipitation rates over a span of decades.   Meteorologists achieve considerable success by using a stochastic framework of analysis  in more or less localized  regions of space-time.

So, a different approach must be taken.   Let us treat the weather system as a deterministic one defined by a set of potentials (of temperature, water concentration, air pressure, etc.) assumed coupled by kinetic interactions which latter we will not attempt to analyze.   For simplicity, because we are dealing with global (as opposed to polar, oceanic, or day-night warming), no spatial factor will be included.

We assume that the kinetics result in a linear (or, stochastically, Markovian) coupling among the potentials, so that system changes do not retain state except in the value of the potentials.   A global Hamiltonian or similar approach would not work, because of complexity in estimating the flux of energy over the long term.

# Entropy$_0$ Decreases

We begin by showing that the entropy (*Entropy$_0$*) associated with the total free energy of the system, formally computed as in thermodynamics, decreases with global warming:

Consider the global temperature $T_t$ as a function of time *t* in increments of a calendar year.  Call the corresponding total system energy $U_t$, and assume it partitioned into kinetic energy $K_t$, potential energy $P_t$, and heat energy $Q_t$.  Based on Hasselmann and others, we consider it established that $T_{2000} - T_{1900}$ amounts to about 0.5° C.  Without affecting the conclusion, we approximate the actual temperatures as $T_{2000} = 290°$ K and $T_{1900} = 289.5°$ K.   We note that a small ~.06 K warming of the oceans has been observed [3] during the latter half of period in question, but we ignore it.

For total energy in the system, we have $U_t = K_t + P_t + Q_t$.  Combining *K* and *P* to represent workable (free) energy *W*, we have

$$U_t = W_t + Q_t \tag{1}$$

Now we define the change in *Entropy$_0$* by the difference, *dS$_t$*,

$$dS_{2000} = dQ_{2000}/T_{2000} - dQ_{1900}/T_{1900} \tag{2}$$

in which *dQ* represents heat flux from the weather.  Avoiding the useless concept of "wasted" heat in an open system, we rewrite (2) using the previous definitions as:

$$dS_{2000} = (U_{2000} - dW_{2000})/T_{2000} - (U_{1900} - dW_{1900})/T_{1900}. \tag{3}$$

We recognize here that the postulated potentials must have different zeroes:  The *U* values represent system totals which are kinetic or aerodynamic transfers to the atmosphere by the Earth's rotation, or are heat or radiative input from the Sun.  So, assuming gas-molecular kinetics or Planckian radiation justifies the use of the Kelvin zero for these potentials.

However, the *W* values primarily represent potentials developed on the atmospheric interaction with water, ice, and land.  Heat of vaporization of water stores about 540 calorie/kg.  Although sea water would freeze below the centigrade zero, liberating about 80 calorie/kg, it would return this free energy at the low end of the potential scale when melted at the centigrade zero.  So, to describe the weather, we consider only the centigrade zero.  The evaluation of *Entropy$_0$* as a term in the overall approximation depends on this approximation, by which we relate the overall zeroes.

Expressing $dW_{1900}$ and $U_{1900}$ in terms of $dW_{2000}$ and $U_{2000}$, we get:

$$dW_{1900} = dW_{2000}(T_{1900}/T_{2000}) \text{ deg C} \cong dW_{2000}(289.5 - 273)/(290 - 273) \quad (4)$$

$$\cong .971\, dW_{2000}; \quad (5)$$

$$U_{1900} = U_{2000}(T_{1900}/T_{2000}) \text{ deg K} \cong U_{2000}(289.5/290) \quad (6)$$

$$\cong .998\, U_{2000}. \quad (7)$$

Substituting (5) and (7) into (3) above for the relation for a negative value of $dS$,

$$dW_{2000} \geq .069\, U_{2000} \quad \text{(about 7\%).} \quad (8)$$

Therefore, the $Entropy_0$ of this system would be expected to have decreased with global warming if the free energy flux exceeded about 7% of the total. A typical value of the free energy flux from the Sun's radiation would be about 15% for evaporative conversion alone [4], so we may be assured of the decrease.

Some comment: For a closed system with a limited store of free energy, as the system did work, the free energy would be seen as being converted irreversibly to heat; the entropy then necessarily would increase until no more work could be done.

During the 19th century, when steam engines were the high technology, there was a theory of the universe that predicted a "heat death": All motion would cease after all the free energy was converted to heat, resulting in a lukewarm, totally disordered mixture, with no potential likely to be found anywhere. This "Big Blah" theory doesn't apply to the open system of the Earth's weather.

## $Entropy_1$ Increases

Next, we show that the second term in our approximation, the $Entropy_1$ of the system, as defined by its randomness but without regard for the total free energy, increases with global warming.

We consider that, knowing the current weather at any given time and place, to the extent one could predict the weather elsewhere (spatially) or into the future (temporally), to that extent would be the $Entropy_1$ of the system lower. In particular, correlation or coherence would imply more organization, more potential for prediction that works, and so more free energy. On the other hand, a high $Entropy_1$ would imply a high amount of unpredictability in the weather in space-time. This definition is consistent with the definition of the $-S p_i \log_2 p_i$ entropy of information theory [5].

Again, looking at the several to perhaps several dozen potentials in the system, we ignore the kinetics and view each potential as being controlled directly by one or more of the others.

Call the potentials $P_i = P_1, P_2, \ldots$, etc. We consider just one potential instance at a time as a representative of any other of the same kind. In general, one of the potentials, $P_i$, will determine another, $P_j$, so that, within small enough intervals,

$$P_j = f(P_i) \cong kP_i \qquad (9)$$

in which $k$ is a constant of proportionality peculiar to the two potentials. We may simplify by treating each $P$ as measured by its absolute value difference from some suitably chosen zero. This is a trivial approach implied directly by the concept of potential. The specific choice of zero may be ignored, because we are considering each potential individually; and, any increase, however coupled, will be an increase away from whatever zero we decide.

Immediately, it may be seen in (9) that a small, possibly random change in $P_i$, $dP_i$, will have an effect proportional to $P_i$. In particular, the standard deviation $\langle dP \rangle_i$ of a potential $P_i$, viewed as a random variable, will be related linearly to that of $P_j$ by the proportionality factor $k$:

$$\langle dP \rangle_j = k \langle dP \rangle_i. \qquad (10)$$

Also, $P_i$ being in the system, an increase in $P_i$ itself will be accompanied by an increase in its standard deviation. This leads directly to the sought result: The standard deviation will increase with an increase in the potential itself. We see that temperature has increased with global warming; therefore, we expect increased variability in the temperature, as well as increased variability in the other potentials in the system, which are coupled to temperature and also may increase in variability for reasons independent of temperature.

## Conclusion

The opposite directions of $Entropy_0$ and $Entropy_1$ are partly because of the merely formal correctness of the thermodynamic definition of $Entropy_0$, and partly because $Entropy_1$, as the second term in an approximate solution to an otherwise intractable problem, implies both spatial and stochastic factors absent from $Entropy_0$. If we look at the meaning of these two terms, we see that, if global warming should continue, the decrease in $Entropy_0$ would mean more free energy to drive the weather; the increase in $Entropy_1$ would mean a harder time predicting it.